\documentclass{article}

\usepackage[final]{neurips_2023_ml4ps}

\usepackage[utf8]{inputenc} 
\usepackage[T1]{fontenc}    
\usepackage{hyperref}       
\usepackage{url}            
\usepackage{booktabs}       
\usepackage{amsfonts}       
\usepackage{nicefrac}       
\usepackage{microtype}      
\usepackage{xcolor}         
\usepackage[none]{hyphenat}
\usepackage{graphicx}
\usepackage{physics}
\usepackage{natbib}
\usepackage{amsmath}
\usepackage{wrapfig}

\title{Unraveling the Mysteries of Galaxy Clusters: Recurrent Inference Deconvolution of X-ray Spectra}

\author{%
Carter Rhea$^{1,2}$ \quad Julie Hlavacek-Larrondo$^{1,3}$ \quad Ralph Kraft$^{4}$ \quad Akos Bogdan$^{4}$   \\
\textbf{Alexandre Adam}$^{1,3,5}$ \quad \textbf{Laurence Perreault-Levasseur}$^{1,3,5,6}$ \\
$^1$Université de Montréal \quad $^2$CRAQ \quad $^3$Ciela Institute \quad $^{4}$ Harvard-Smithsonian CfA \\ \  $^5$Mila \quad $^6$CCA, Flatiron Institute\\ 
\texttt{\{carter.rhea,j.larrondo,alexandre.adam,}\\ \texttt{laurence.perreault.levasseur\}@umontreal.ca}\\
\texttt{\{rkraft,abogdan\}@cfa.harvard.edu}
\\
}

\begin{document}

\maketitle

\begin{abstract}
  In the realm of X-ray spectral analysis, the true nature of spectra has remained elusive, as observed spectra have long been the outcome of convolution between instrumental response functions and intrinsic spectra. In this study, we employ a recurrent neural network framework, the Recurrent Inference Machine (RIM), to achieve the high-precision deconvolution of intrinsic spectra from instrumental response functions. Our RIM model is meticulously trained on cutting-edge thermodynamic models and authentic response matrices sourced from the \textit{Chandra} X-ray Observatory archive. Demonstrating remarkable accuracy, our model successfully reconstructs intrinsic spectra well below the 1-$\sigma$ error level. We showcase the practical application of this novel approach through real \textit{Chandra} observations of the galaxy cluster \textbf{\textit{Abell 1550}}—a vital calibration target for the recently launched X-ray telescope, \textit{XRISM}. This work marks a significant stride in the domain of X-ray spectral analysis, offering a promising avenue for unlocking hitherto concealed insights into spectra.

\end{abstract}

\section{Introduction}
We describe a methodology to deconvolve intrinsic X-ray spectra from the instrumental response using machine learning; we apply this to the case of galaxy clusters.
Galaxy clusters harbor a  large reservoir of hot gas ($\sim 10^7-10^8$), called the IntraCluster Medium (ICM), which accounts for the majority of baryonic matter in the cluster (\textit{e.g.}  \citealt{fabian_x-rays_2003}). This gas consists primarily of ionized hydrogen and helium but also contains numerous heavier elements (\textit{e.g.}  \citealt{mushotzky_x-ray_1984}; \citealt{mohr_properties_1999}; \citealt{loewenstein_chemical_2003}). Emission mechanisms such as thermal bremsstrahlung, bound-free atomic transitions, and the collisional excitation of hydrogen are responsible for the X-ray continuum (\textit{e.g.}  \citealt{markevitch_comparison_1997};  \citealt{ettori_coulomb_1998};  \citealt{sarazin_x-ray_1999};  \citealt{markevitch_temperature_1998}). 
The collisionally-ionized gas also exhibits strong emission lines in its spectra coming from the excitation of heavy elements such as nickel and iron.

Observed galaxy cluster spectra in the X-ray regime, $S(E')$, are the result of an integration between the intrinsic spectrum of the cluster, $F(E)$, and the instrumental response, $R(E',E)$:
$
    S(E') = \int_0^\infty R(E',E)F(E)dE +\eta, 
$
where $E'$ is the measured photon energy,  $E$ is the true photon energy, and $\eta$ is the noise generally modeled by a Poisson distribution \texttt{Pois}$(\lambda$). In this  work, we only consider high signal-to-noise (SNR$>$50) observations where the noise approaches a Gaussian distribution described as $\mathcal{N}(0, \sigma^2\mathcal{I})$. Since we discretely sample the energies, the functional form is reduced to a matrix form, reducing the equation to 
\begin{equation}\label{eqn:forward}
\textbf{S}_i = \sum_{j}R_{ij}\textbf{F}_j + \boldsymbol{\eta}_i.
\end{equation}
Finally, 
Despite the simplicity of this linear equation, the response matrix, $R_{i,j}$, is highly singular, thus rendering a standard inversion impossible for complex models. \cite{rhea_data_2021} demonstrated that even sophisticated regularization techniques such as the Moore-Penrose pseudo inverse and Tikhonov regularization fail to resolve the issue.

In the last several years, a new machine learning algorithm known as a recurrent inference machine (RIM) has been developed and shown to excel at solving inverse problems (\citealt{Putzky_recurrent_2017}). The RIM has been used extensively to solve inverse problems such as deconvolution and denoising (see for example \citealt{Morningstar_Data_2019}; \citealt{Morningstar_Analyzing_2018};\citealt{2021arXiv210412864M}; \citealt{2022mla..confE..48A}; \citealt{adam_rim_2023}).  In the context of astrophysics, the RIM has been used to recover undistorted images of background sources in gravitational lenses outperforming standard techniques considerably. In a similar fashion, we use a RIM to solve our inverse problem, equation \ref{eqn:forward}, and thus uncover the intrinsic spectrum of an X-ray source. Machine learning has been previously used to solve similar deconvolution problems in astronomy (e.g. \citealt{molnar_spectral_2020}).
We demonstrate the use of this algorithm on data from the \textit{Chandra X-ray Observatory} -- one of the world's leading X-ray telescopes. 

While in this initial study, we demonstrate the RIM deconvolution method, once deconvolved, multiple science applications for the intrinsic spectra become possible. For example, they could be used as inputs to a convolutional neural network to predict point estimates of the underlying thermodynamic parameters. With these estimates, we can apply the technique described in \cite{legin_beyond_2023}, which uses these quantities in a Simulation-Based Inference \citep{doi:10.1073/pnas.1912789117} framework to recover posteriors on these parameters. 
We can also use the deconvolved spectra to test the line-by-line calibration of X-ray telescopes since, for a calibration target, the source's intrinsic spectrum is non-changing. Therefore, any observed change in the deconvolved spectrum is a result of errors in the response.

\section{Data}
\begin{figure}\label{fig:Conv}
  \centering
  \includegraphics[width=1\textwidth]{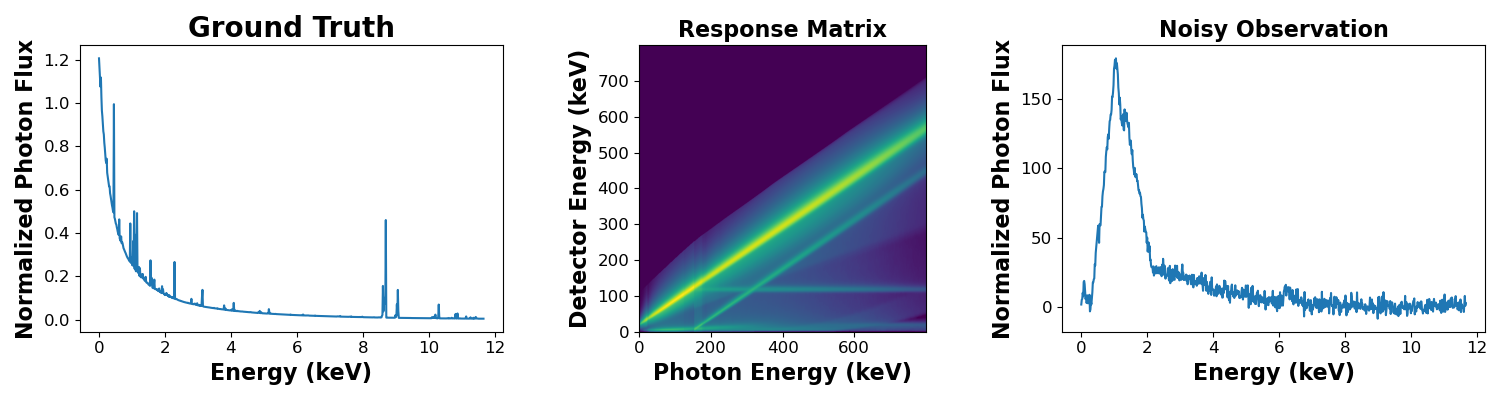}
  \caption{Schematic of the convolution applied by the ACIS instrument on \textit{Chandra}. This graphic demonstrates the response matrix's profound effect on the observed spectrum. The true spectrum is a typical emission spectrum from the ICM modeled using the \texttt{apec} model with a temperature of 2.0 keV and a metallicity of 0.3 Z$_\odot$. The response matrix was taken from a randomly chosen ObsID (2707); the matrix displayed is the log of the response matrix. The noisy observation is created by convolving the response matrix with the ground truth spectrum and then adding noise at a signal-to-noise level of 50.}
\end{figure}

We construct 50,000 synthetic X-ray spectra using state-of-the-art thermodynamic models implemented in \texttt{SOXS} -- a sophisticated modeling pipeline developed for X-ray observatories. We sample widely over the parameters that influence the spectra.
We use the \texttt{APEC} model, which simulates the emission spectrum of a collisionally-ionized diffuse gas using a sophisticated database of atomic transition lines, AtomDB (\citealt{smith_collisional_2001}; \citealt{foster_updated_2012}). 
There are four primary variables that dictate the shape of \texttt{APEC} spectra: the temperature of the gas, the relative metalicity of the gas (i.e., what is the relative abundance of different metals -- elements other than hydrogen and helium), the redshift of the gas which corresponds to a shift in the observed wavelength of the emission lines, and the normalization parameter which described the normalization of the model to the data.
We randomly selected from a uniform distribution of each parameter to construct our synthetic spectra; the values were chosen to adhere to the values we find in galaxy clusters (e.g., \citealt{mcdonald_remarkable_2017}; \citealt{fabian_observational_2012}; \citealt{markevitch_lx-t_1998}). The temperature was set between 0.5 keV (kilo-electron Volts) and 8.0 keV, corresponding to approximately $10^7$K to $10^8$K. The metalicity was sampled between 0.2 $Z_\odot$ (solar metallicity) to 1.2 $Z_\odot$. Since we wish to use this methodology to study nearby (i.e., low-redshift) objects, we allow the redshift to vary between 0 and 0.25. 
In figure \ref{fig:Conv}, we show an example intrinsic spectrum (left), an example response matrix (center), and an example observation (right) created from the convolution between the response matrix and the intrinsic spectrum plus noise.

We mined the \textit{Chandra X-ray Observatory} archive\footnote{https://cda.harvard.edu/chaser/} to construct a set of 1000 response matrices. The observations from which the response matrices were sampled were known galaxy clusters with long exposures (\citealt{mushotzky_x-ray_1984}). However, the response matrix does not depend on the object observed.
Since $R$ is the critical variable in our function and can change as a function of time and position on the detector, sampling an ample amount of real data is critical.
Using our forward model, equation \ref{eqn:forward}, we are able to construct mock observed spectra quickly. Moreover, we randomly construct the noise profile, $\eta$,
where $\eta$ is drawn from a Gaussian distribution representing a signal-to-noise of 30 to 100. In making this choice, we restrict our models to only high-count statistics.

We test our trained network on real observations by the \textit{Chandra X-ray Observatory}. We select \textbf{\href{http://cdsportal.u-strasbg.fr/?target=NGC\%201550}{\color{blue} {\textit{NGC 1550}}}} a dynamically stable galaxy cluster with a well-constrained global temperature and metallicity, since it is the calibration target for the newly launched X-ray observatory \textit{XRISM} (\citealt{xrism_science_team_science_2020}). This target was chosen because previous spectral analysis indicates that a single thermal model is required to model the ICM emission fully (\citealt{kolokythas_evidence_2020}). 
After downloading the raw observation data from the Chandra Archive (ObsID 3186 \& ObsID 3187), we clean the data using the \texttt{chandra\_repro} (v.4.15) pipeline provided by the \textit{Chandra X-ray Center} software group. The spectrum is extracted by taking a large circular region centered on the centroid of the X-ray emission, reaching an average signal-to-noise of 50. During the spectral extraction process, we also calculate the region's response matrix and the ObsID). 

\section{Methods}

Our goal is to recover an estimate of the true spectrum $F \in \mathbb{R}_+^n$ of a galaxy cluster given an observed spectrum $S \in \mathbb{R}_+^m$ from the \textit{Chandra X-ray Observatory}. The observed spectrum is related to the intrinsic spectrum by a linear response function $R \in \mathbb{R}^{m \times n}$ and additive noise $S = RF + \eta$. Due to the ill-posedness of this problem, we must introduce a regularization term, which is most naturally viewed in a Bayesian settings as a prior distribution over the intrinsic flux $p(F)$. The maximum a posteriori (MAP) solution maximizes the product of the likelihood $p(S \mid F)$ and the prior, which can be written as follows
\begin{equation}\label{eq:MAP}
    \hat{F}_{\mathrm{MAP}} = \underset{F}{\mathrm{arg}\,\mathrm{max}}\, 
            \log p(S \mid F) + \log p(F)
\end{equation}
To solve this problem, we make use of a Recurrent Inference Machine (RIM) \citep{Putzky_recurrent_2017}. In this meta-learning framework, the prior or the regularization term is learned implicitly in a neural network $g_\theta$ with parameters $\theta$, which solves equation \eqref{eq:MAP} iteratively via a recurrent series akin to gradient ascent
\begin{equation}\label{eq:recurrent series}
\begin{aligned}
    F_{t+1} &= F_t + g_\theta(F_t, \grad_{F}\log p(S \mid F), \mathbf{h}_t) \\
    \mathbf{h}_{t+1} &= g_\theta(F_t, \grad_{F}\log p(S \mid F), \mathbf{h}_t)\,
\end{aligned}
\end{equation}
where $t \in \{0,\dots, T -1\}$ is the step index and $\mathbf{h}_t$ is an hidden state.
Since the likelihood is Gaussian, we can readily evaluate the likelihood for any trial point $F_t$ using the formula
\begin{equation}\label{eqn:likelihood}
    \grad_{F_t} \log p(S \mid F_t) = (S - R F_t)^TC^{-1}R
\end{equation}
where $C \in \mathbb{R}^{m \times m}$ is the covariance of the additive Gaussian noise of a particular observation. 

At each step of the recursion, the input of the neural network is the gradient of the log-likelihood and the current best reconstruction. The architecture, based on previous works \cite{Morningstar_Data_2019}, is a Convolutional Neural Network (CNN) augmented with Gated Recurrent Units (GRU; \citealt{chung_empirical_2014}) to model the dynamical evolution of the recurrent series \eqref{eq:recurrent series}.
After this, the output is the reconstructed intrinsic spectrum, which we feed to the forward model in equation \ref{eqn:forward} to produce a model, which we then use to evaluate the gradient of the likelihood wrt the solution (equation \ref{eqn:likelihood}). This output is then fed back in the network at the next step of the recursion.
We use a mean squared loss function:
\begin{equation}
    \mathcal{L} = \frac{1}{T}\sum_{t=1}^T\sum_{i=1}^M (\hat{x}_i^{(t)} - x_i)^2
\end{equation}
where $\hat{x}_i^{(t)}$ is the current best reconstruction at time $t$ and $M$ is the total number of spectral channels in the spectrum.


\section{Results and Discussion}
To asses the performances of the RIM on unseen data, we build a test set containing 1,000 new examples containing 100 new response matrices and 100 new simulated spectra.

\begin{figure}
    \centering
    \includegraphics[width=0.998\textwidth]{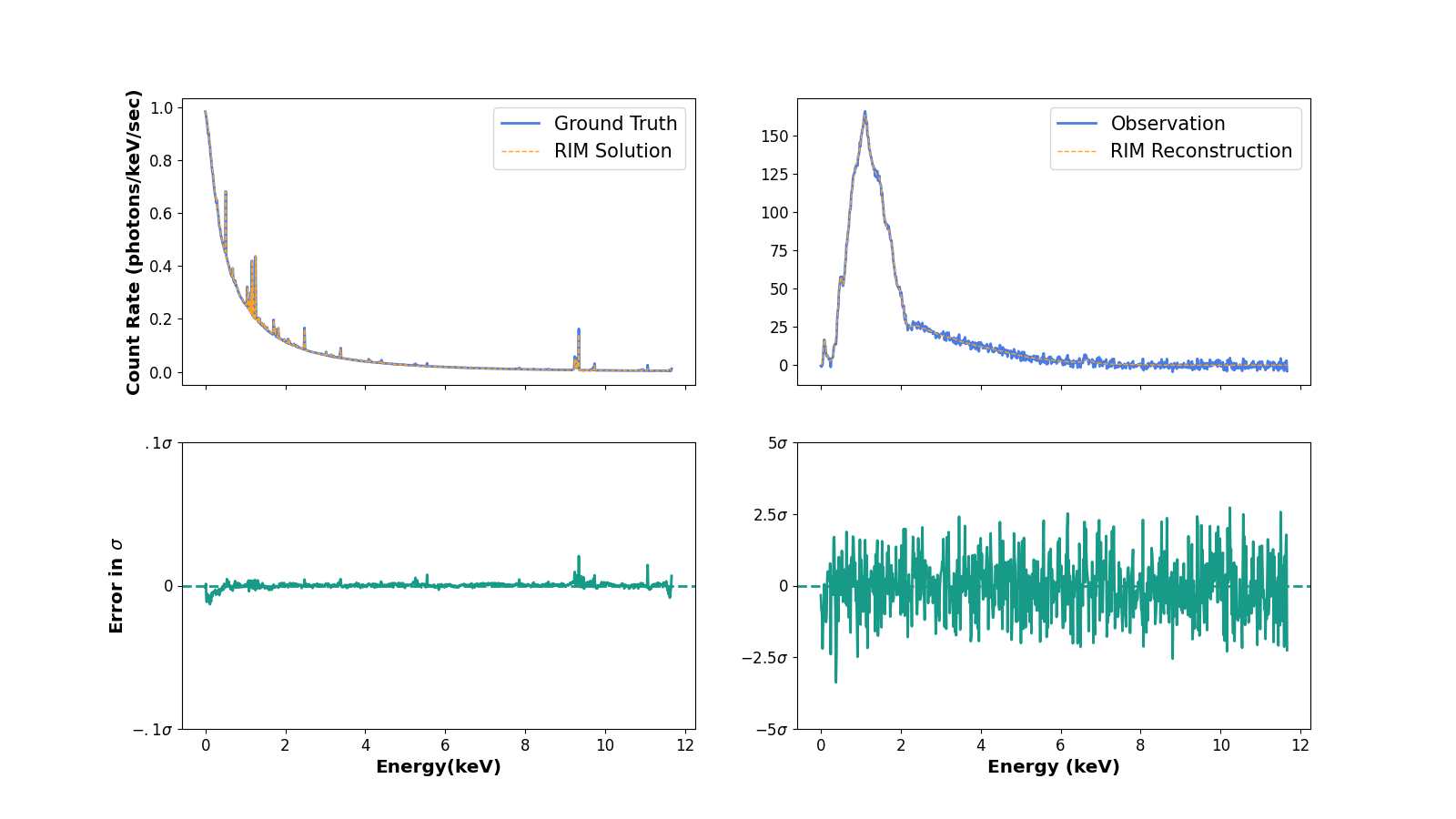}
    \caption{This graphic shows the results of the trained RIM on a randomly selected spectrum from the test set. The top left panel shows the true intrinsic spectrum (solid blue) and the RIM solution (dashed orange). Below, we report the residual between the true solution and the RIM solution normalized by the noise level of the observed spectrum. In the top right panel, we show the observed spectrum (solid blue) and the result of passing the RIM solution through the forward model (dashed orange). Below, we report the residual between the two curves normalized by the observed noise level.}
    \label{fig:RIMsynthetic}
\end{figure}

In figure \ref{fig:RIMsynthetic}, we show the results of the trained RIM on a randomly selected spectrum from the test set of synthetic data. The graphic demonstrates that the RIM has accurately learned how to reconstruct the underlying spectrum from just the observed spectrum through the gradient of the likelihood and response matrix.
The reconstructed intrinsic spectrum (dashed purple) fits the intrinsic spectrum (solid blue) to less than 0.1$\sigma$ error. The reconstructed intrinsic spectrum captures all the emission lines in the true spectrum. On the right panel, we compare the observed spectrum  and the modeled convolved spectrum obtained by feeding the RIM solution through the forward model (without adding noise). We display the residual which is the difference between the ground truth and the RIM prediction. These residuals are consistent with noise, as they should for an accurate solution to the inverse problem (see the bottom right panel of figure \ref{fig:RIMsynthetic}). 
these features disappear. Therefore, the RIM effectively denoises the spectrum.

\begin{figure}
    \centering
\includegraphics[width=0.995\textwidth]{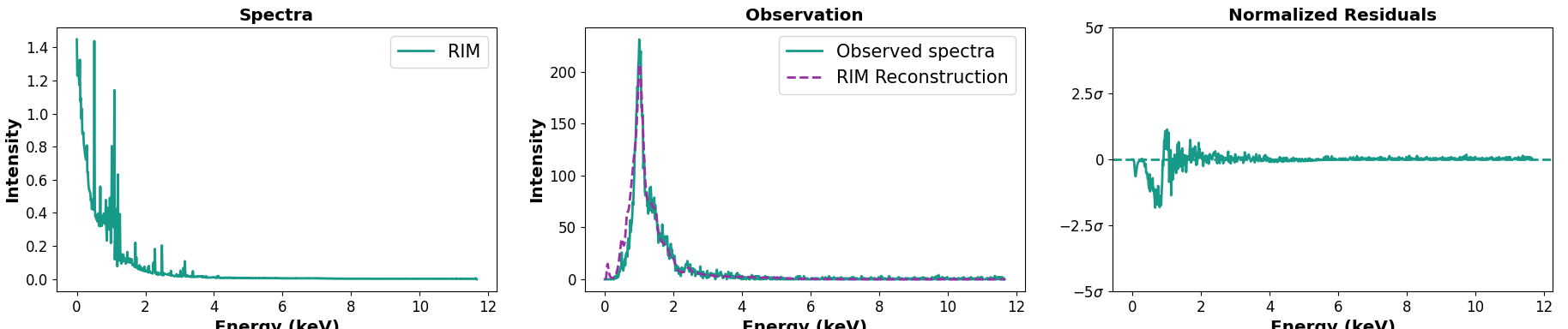}
    \caption{On the left, we show the RIM recovered spectrum of NGC 1550 from ObsID 3186. In the middle, we compare the observed spectrum of NGC 1550 (solid teal) to the RIM-reconstructed spectrum that is created by applying the forward model to the RIM-recovered spectrum (dashed purple). The reconstruction is in strong agreement with the observed spectrum, achieving a less than 1-$\sigma$ disagreement between the noisy observed spectrum and the RIM reconstructed observation. 
    A valid solution to the inverse problem should leave residuals consistent with noise.
    On the right, we show the residual between the observed spectrum and the result of the forward model of the RIM reconstruction normalized by the noise level.}
    \label{fig:NGC1550}
\end{figure}

Figure \ref{fig:NGC1550} depicts the RIM reconstructed intrinsic spectrum of the central ICM emission of \textbf{\textit{NGC 1550}} (ObsID 3186) on the left panel, while, on the right panel, we plot the noisy observed spectra (solid teal) and the RIM reconstructed spectrum (dashed purple). As expected, the recovered intrinsic spectrum exhibits strong emission lines around 1 keV and an underlying power-law continuum. Moreover, the right panel demonstrates that, when used in the forward model, the RIM reconstructed spectrum matches the observed spectra below the 1-$\sigma$ level in the busiest part of the spectrum (between 0.5 and 2 keV) and well below the 0.1-$\sigma$ level throughout the rest of the spectrum. Although we only show the results for one ObsID above, we obtain similar results using the other \textit{Chandra} observation ObsID 3187.
 
\section{Summary and Future Improvements}
In conclusion, we use a recurrent inference machine to deconvolve X-ray spectra of galaxy clusters using only the observed spectrum and the associated response matrix. Our network achieves results consistent with the noise statistics on our quasi-synthetic dataset constructed using state-of-the-art thermodynamic models and real response matrices mined from the \textit{Chandra} archive. Crucially, when applied to the case of a high signal-to-noise observation of a galaxy cluster, \textbf{\textit{Abell 1550}}, the RIM is able to uncover the intrinsic spectrum; the RIM-reconstructed observation matches perfectly to the noise level of the observation. This work serves as a proof-of-concept for a powerful novel paradigm in X-ray spectral analysis.

Future work will consist of building this method into existing spectral fitting pipelines to improve fit results and applying it to more high signal-to-noise observations of galaxy clusters. We also will extend the suite of synthetic data to include more complex multi-temperature models in order to extend this application to more complicated galaxy clusters. Additionally, this methodology is not limited to the \textit{Chandra X-ray Observatory} but can be used on any X-ray telescope if retrained using its response matrices; we will be applying this method to recover the intrinsic spectra of \textit{XRISM} targets once its calibration is complete. This method opens up X-ray spectra to other deep learning techniques to recover the thermodynamic parameters of the intrinsic source.

\begin{ack}
The work is in part supported by computational resources provided by Calcul Quebec and the Digital Research Alliance of Canada. The work of A.A. was partially funded by a National Sciences and Engineering Council of Canada (NSERC) CGS D scholarship. J.H.L and L.P.L. acknowledges support from the Canada Research Chairs Program, NSERC through grant RGPIN-2020-05102, and the Fonds de recherche du Québec through grant 2022-NC-300397.

\end{ack}


{
\small
\bibliography{CNN-Xray}
}

\end{document}